\documentstyle{article}
\title{
{\bf THE SPECTRUM OF THE ANOMALOUS DIMENSIONS \\
OF THE COMPOSITE OPERATORS \\
IN  $\epsilon$-EXPANSION \\
IN THE SCALAR $\phi^4 $- FIELD THEORY}
}
\author{S. \'{ E}. Derkachov\\
Department of Mathematics, \\
St.-Petersburg Technology Institute\\
and\\
A. N. Manashov\\
Department of Theoretical Physics,\\  Sankt-Petersburg State
University \\ }

\begin{document}

\large
\maketitle

\begin{abstract}
{The spectrum of the anomalous  dimensions of the composite operators
(with arbitrary number of fields $n$ and derivatives $l$)
in the scalar $\phi^4$ - theory in the first order of the
$\epsilon$ -expansion  is investigated. The exact solution for the
operators with number of fields $\leq 4$ is presented.
The behaviour of the anomalous dimensions in the large $l$ limit
has been analyzed. It is given the qualitative description of the
structure of the spectrum  for  the arbitrary $n$.}
\end{abstract}
\newpage

\renewcommand{\theequation}{\thesection.\arabic{equation}}

 \section{Introduction}

The subject of this paper is to study the critical
scalar $\phi^4$ - model in first order in
$\epsilon$ - expansion.
In recent papers \cite{1,2} the spectrum of the
anomalous dimensions of the composite operators
have been investigated in N-vector model
(first order in $\epsilon$).
Due to mixing of the operators the
general problem is not tractable
already in one-loop level.
In \cite{1,2} the solution has been obtained  for the operators
with two and three fields where this difficulty
does not  present in full.
In our view main problem appeared already in scalar theory is the
mixing of the operators,
the "N-vector complication" being not
principial. (Note, that the problem of calculation of the spectrum
of the anomalous dimensions for the composite operators in scalar
$\phi^4$ theory  is equivalent those for $O(N)$ symmetric and
traceless ones in N-vector theory.)
By this reason we confine ourselves by consideration of spatially
symmetric and traceless composite operators in scalar $\phi^4$ -
theory.  Using approach different from one in \cite{1,2} we
succeeded in obtaining the exact solution for operators with four
fields. Some properties of the latter  is appeared to be inherent
in spectrum for arbitrary $n$. (We prove that all nonzero
eigenvalues of the spectrum for given $n$ are appeared as either
the accumulating points or  the exact eigenvalues of infinite
degeneracy in the spectrum on $(n+k)$ level.)

The organisation of this
paper is as follows.
In Sect.\ref{descr} we give the full description of the
canonical conformal  spatially
symmetric and traceless composite
operators. This extend the well know
solution for the case of two fields \cite{YM} and give
in this special case the same result.
The similar problem has been considered
recently by  F.Wegner and S.Kehrein \cite{2}.
In Sect.\ref{oneloop} we
calculate the counterterms in first
order in  coupling constant and formulate
main equations.
These equations were obtained in \cite{1,2} in a different
manner.
In Sect.\ref{dim23} it is shown how
the known solutions of the eigenvalue
problem for the operators with two \cite{WK} and three \cite{2}
fields can be  obtained  in the frame of our approach.
Sec.\ref{four} is devoted to the investigation of the eigenvalue
problem for the operators with four fields.
In Sec.\ref{tree} we prove some statement about structure of
spectrum valid in general case and discuss the qualitative picture of
the distribution of the eigenvalues.

\setcounter{equation}{0}
\section{Spatially symmetric and traceless
composite operators.}
\label{descr}

The composite operator is a local functional of the
field $\phi(x)$ which contains $n$ fields and $l$
spatial derivatives
acting on these fields. We suppose that all indices
of the derivatives are "free"
i.e. operator have as many tensor indices as a derivatives.
We consider only the spatially symmetric and traceless
composite operators $\Psi_{i_1i_2...i_l}(x)$.

It is useful to introduce the "scalar" operators
$$
\Psi(x,u) \equiv \Psi_{i_1...i_l} u_{i_1}...u_{i_l}
$$
where $u_i$ is the D-dimensional vector.
For the sake of brevity we shall call  $\Psi(x,u)$ by  composite
operator also. It is easy to understand that
every composite operator $\Psi(x)$  with n fields
is defined by the total symmetric
"coefficient functions" of n arguments
$\psi(z_1,...,z_n)$ in the following way:
\begin{equation}
\Psi(x,u) = \left.
\psi(\frac{\partial}{\partial a_1},...,
\frac{\partial}{\partial a_n})
\Phi(x;a_1,...,a_n) \right|_{a_m = 0} ;
\label{def1}
\end{equation}
where
\begin{equation}
\Phi(x;a_1,...,a_n) \equiv
\left. P(u , \frac{\partial}{\partial v})
\prod_{m=1}^n \phi(x+a_m v)\right|_{v=0}.
\label{def2}
\end{equation}
$P$ -  projector on subspace of traceless tensors.
$$
\frac{\partial^2}{\partial u_i \partial u_i} P(u , v) =
\frac{\partial^2}{\partial v_i \partial v_i} P(u , v) = 0
\ , \  \left. P(u , \frac{\partial}{\partial w}) P(w , v)
\right|_{w=0} = P(u , v).
$$
The explicit expression for function
 $P(u , v)$
have form \cite{PD}:
$$
P(u , v) = F(uv , u^2v^2 ) \ , \  \mbox{{and at}} \   {x^2 \leq y}
$$
$$
F(x , y) = \frac{\Gamma(\mu-1)}{2\pi i}
\int \frac{(4z)^{\mu-1}e^{z}{\rm d}z}
{(y - 4zx + 4z^2)^{\mu-1}},
$$
here $\mu=(D-2)/2$, and $D$ -- the dimension of space.
The integrand function is defined on the complex
$z$-plane
with three horizontal cuts, going from the points
$z_1=0, \> z_2=(x+\sqrt{x^2-y}), \> z_3=(x-\sqrt{x^2-y})$ to $
 -\infty$.
The integration contour is shown on fig.1.
 \vspace{0.5cm}

\unitlength=1mm
\special{em:linewidth 0.4pt}
\linethickness{0.4pt}
\begin{picture}(137.00,77.00)
\put(107.00,43.00){\oval(16.00,36.00)[r]}
\put(107.00,61.00){\line(-1,0){95.00}}
\put(107.00,25.00){\line(-1,0){95.00}}
\put(41.00,61.00){\vector(1,0){2.00}}
\put(38.00,25.00){\vector(-1,0){2.00}}
\put(55.00,43.00){\circle*{2.00}}
\put(104.00,57.00){\circle*{2.00}}
\put(104.00,28.00){\circle*{2.00}}
\put(104.00,28.00){\line(-1,0){93.00}}
\put(104.00,57.00){\line(-1,0){93.00}}
\put(55.00,43.00){\line(-1,0){44.00}}
\put(55.00,43.00){\line(1,0){82.00}}
\put(55.00,77.00){\line(0,-1){64.00}}
\put(60.00,46.00){\makebox(0,0)[cc]{z = 0}}
\put(87.00,52.00){\makebox(0,0)[cc]{$z =2 x + 2\sqrt{x^2 - y}$}}
\put(87.00,32.00){\makebox(0,0)[cc]{$z = 2 x - 2\sqrt{x^2-y}$}}
\put(55.00,7.00){\makebox(0,0)[cc]{(Fig.1}}
\end{picture}

In the case when operators have as many tensor indices
as a derivatives, function
$\psi(z_1,...,z_n)$ is the gomogeneous symmetric polynomial
of its n arguments.

A basis for the vector space of symmetric, gomogeneous
polynomials of degree l is generated by product
$$
[s_1(z_1,...,z_n)]^{m_1}
[s_2(z_1,...,z_n)]^{m_2}[s_3(z_1,...,z_n)]^{m_3}...
[s_n(z_1,...,z_n)]^{m_n},
$$
where $ m_1 + 2m_2 + 3m_3 +...+n m_n = l$ and $m_i \geq 0$.
The polynomials $s_n(z_1,...,z_n)$ are the standard symmetric
polynomials.
The generating function for these polynomials is
$$
\prod_{m=1}^n (z_m + t) =
\sum_{m=1}^n t^{n-m} p_n(z_1,...,z_n).
$$

 All composite operators are divided
in two following sets. The operators from first set may be
represented as a $\partial_i...\partial_k \Psi_{s_1...s_m}$
and for operators from second set it is impossible
(nonderivative operators).
The operators $\partial_i...\partial_k \Psi_{s_1...s_m}$
and $\Psi_{s_1...s_m}$ have the equal anomalous dimensions.

The operators from first set have the
"coefficient functions" $\psi(z_1,...,z_n)$ of the
following structure
$$
\psi(z_1,...,z_n) = (z_1+z_2+...+z_n)^p \phi(z_1,...,z_n).
$$

To extract all these operators,
it is convenient to introduce the restriction
$$
z_1 + z_2 + ...+ z_n = 0
\Longleftrightarrow s_1(z_1,...,z_n) = 0
$$
A basis for the vector space of symmetric , gomogeneous
polynomials of degree l with this restriction
is generated by product
$$
[s_2(z_1,...,z_n)]^{m_2}[s_3(z_1,...,z_n)]^{m_3}...
[s_n(z_1,...,z_n)]^{m_n}
$$
where $2m_2 + 3m_3 +...+n m_n = l$ and $m_i \geq 0$
(and we keep in mind the restriction $s_1 = 0$).

 In Sec.\ref{dim23} it will be shown that to obtain the spectrum
 of the anomalous dimension is enough to know the block of the mixing
 matrix describing the renormalization on the conformal operators.
So we give the full description of the conformal operators.

Let $x \rightarrow x^{\prime}(x)$ be the coordinate
transformation of the general form  and $\phi \rightarrow
\phi^{\prime}$ is the special field transformation of the form :
\begin{equation}
\phi(x) \rightarrow \phi^{\prime}(x) =
{\rm det}{\left [\frac{\partial x^{\prime}}{\partial
x}\right ]}^{\Delta / D} \phi(x^{\prime}) .  \label{trans}
\end{equation}
Here $\Delta$ is a constant which is called the
canonical field dimension (for the $\phi^4$-model $\Delta =
(D-2)/2$).
For the infinitesimal transformations~(\ref{trans}) one obtains
$$
x \rightarrow x^{\prime}(x)= x+\alpha(x)
$$
\begin{equation}
\phi(x) \rightarrow
\phi^{\prime}(x)= \phi(x)+\delta_{\alpha}\phi(x) \ ,\
\delta_{\alpha}\phi(x) \equiv
(\alpha(x)\partial + \frac{\Delta}{D}\partial \alpha (x))\phi(x)
\label{tr1}
\end{equation}

The transformations (\ref{tr1})
  generate the following
transformations of the composite operators
\begin{equation}
\delta_{\alpha}\Psi_{i_1i_2...i_l}(x) \equiv
\frac{\delta \Psi_{i_1i_2...i_l}(x)}{\delta \phi}
\delta_{\alpha}\phi, \> \>  \mbox{{\rm or}} \> \> \>
\delta_{\alpha}\Psi(x,u) \equiv
\frac{\delta \Psi(x,u)}{\delta \phi}
\delta_{\alpha}\phi.
\label{indus}
\end{equation}
For the infinitesimal conformal transformations one has:
$$
\alpha_i(x) = a_i + \omega_{ik}x_k + \lambda x_i +
(x^2\delta_{ik} - 2x_ix_k)b_k
$$
where $a_i , \omega_{ik}=-\omega_{ki} , \lambda , b_k $ are
(constant) group parameters.
Summation over repeated indices will be understood throughout.

The composite operator is called canonically conformal
one if the transformation law of this operator
 for conformal $\alpha$
 has the form:
\begin{equation}
\delta_{\alpha}\Psi_{i_1...i_l} =
(\alpha\partial + \frac{\Delta_{\Psi}}{D}
\partial \alpha )\Psi_{i_1...i_l} +
\sum_{k=1}^l \partial_{i_k} \alpha_s
\Psi_{i_1...s...i_l}
\end{equation}
 or in equivalent form,
\begin{equation}
\delta_{\alpha}\Psi(x,u) =
(\alpha\partial + \frac{\Delta_{\Psi}}{D}
\partial \alpha )\Psi(x,u) +
((u\cdot \partial) (\alpha\cdot\partial_u))
\Psi(x,u).
\label{ind1}
\end{equation}
(Note, that $\Delta_\psi=$ (scale dimension)-(number of indices)).

The requirement of the conformal invariance
of the $\Psi(x;u)$ leads to the following
expression for
the canonical dimension $\Delta_{\Psi} = n \Delta$
and to the differential equation for the symmetric and
gomogeneous function $\psi(z_1,...,z_n)$ (see Appendix A):
\begin{equation}
\sum_{m=1}^n (z_m \frac{\partial^2}{\partial z_m^2} +
\Delta \frac{\partial}{\partial z_m}) \psi(z_1,...,z_n) = 0.
\label{4f}
\end{equation}

In the $\phi^4$ theory  $\Delta = (D-2)/2$
and $D = 4 - 2 \epsilon$. For $\epsilon = 0$ $\Delta = 1$ and
therefore eq.~(\ref{4f}) turns into
\begin{equation}
\label{3a} \sum_{m=1}^n (z_m
\frac{\partial^2}{\partial z_m^2} + \frac{\partial}{\partial
z_m}) \psi(z_1,...,z_n) = 0.
\end{equation}
Let us consider
 the following transformation of $\psi \rightarrow \hat{\psi}$
\begin{equation}
\label{5a}
\hat{\psi}(z_1,...,z_n) = \int_{0}^{\infty}
\prod_{m=1}^n {\rm d}t_m {\rm e}^{-t_m}
\psi(t_1 z_1 ,...,t_n z_n) .
\end{equation}
It is not difficult to obtain the
explicit form of this transformation.
If the function $\psi(z_1,...,z_n)$ has the form
$$
\psi(z_1,...,z_n) = \sum_{(i)}
\psi_{m_1...m_n} z_1^{m_1}...z_n^{m_n}
$$
then for the function
$\hat{\psi}(z_1,...,z_n)$ we obtain
$$
\hat{\psi}(z_1,...,z_n) = \sum_{(i)}
\psi_{m_1...m_n} m_1!...m_n!  z_1^{m_1}...z_n^{m_n}
$$
The equation~(\ref{3a}) in the terms of $\hat{\psi}$ takes
 the more simple form:
\begin{equation}
\label{9a}
\sum_{m=1}^n \frac{\partial}{\partial z_m}
\hat{\psi}(z_1,...,z_n) = 0.
\end{equation}
For some purpose it is necessary to generalize
this consideration.
Let us suppose that composite operator is builded from
a different "elementary fields":
\begin{equation}
\Psi(x,u) = \left.
\psi(\frac{\partial}{\partial a_1},...,
\frac{\partial}{\partial a_n})
\prod_{i=1}^n \phi_i(x + a_i u) \right|_{a_m = 0}
\end{equation}
and each $\phi_i$ has the simple transformation
law with respect to the general coordinate
transformation:
$$
\delta_{\alpha}\phi_i(x) =
(\alpha(x)\partial +
\frac{\Delta_i}{D}\partial \alpha (x))\phi(x).
$$
The requirement of the conformal invariance
of the $\Psi(x;u)$ leads to the equation
on the function $\psi(z_1,...,z_n)$
(now this function is not symmetric):
\begin{equation}
\label{333a}
\sum_{m=1}^n (z_m \frac{\partial^2}{\partial z_m^2} +
\Delta_m \frac{\partial}{\partial z_m}) \psi(z_1,...,z_n) = 0.
\end{equation}
Let $\psi \rightarrow \hat{\psi}$ be the following
transformation
\begin{equation}
\label{5ab}
\hat{\psi}(z_1,...,z_n) = \int_{0}^{\infty}
\prod_{m=1}^n {\rm d}t_m \ t_m^{\Delta_m - 1}{\rm e}^{-t_m}
\psi(t_1 z_1 ,...,t_n z_n) .
\end{equation}
The equation~(\ref{333a}) on the function $\psi$ also
turns into the simple one
on the function $\hat{\psi}$
\begin{equation}
\sum_{m=1}^n \frac{\partial}{\partial z_m}
\hat{\psi}(z_1,...,z_n) = 0.
\end{equation}
This equation show that function $\hat{\psi}$ is
translation invariant and the solutions of this equation
are described in \cite{2}.

Taking into account eq.~(\ref{9a}) we conclude that the space of
the conformal  operators is equivalent to
 the vector space $C(n,l)$ of symmetric , gomogeneous
and translation invariant polynomials of n variables.
A basis for it is generated by product
$$
[p_2(z_1,...,z_n)]^{m_2}[p_3(z_1,...,z_n)]^{m_3}...
[p_n(z_1,...,z_n)]^{m_n},
$$
where $2m_2 + 3m_3 +...+n m_n = l$ and $m_i \geq 0$.
The polynomials $p_i(z_1,...,z_n)$ can be obtained from
the following formula:
$$
\prod_{m=1}^n (z_m - \frac{1}{n}\sum_{k=1}^n z_k + t) =
\sum_{m=1}^n t^{n-m} p_m(z_1,...,z_n).
$$
The generating function $\rm{D}(n,x)$ for
 $\rm{dim} C(n,l)$ has
the form:
\begin{equation}
\label{77}
\rm{D}(n,x) \equiv
\sum_{l=0}^{\infty} \rm{dim}C(n,l)\> x^l =
\frac{1}{\prod_{i=2}^{n}(1-x^i)}
\end{equation}

\setcounter{equation}{0}
\section{The one-loop counterterms.}
\label{oneloop}

Let us consider the renormalization of
the set of the operators  described above.
The renormalized operator is defined as (\cite{AN,Col}):
\begin{equation}
[\Psi_i (\phi)]_R =
\sum_{k} Q_{ik} \Psi_k (\phi) \ ,
\end{equation}

where $Q_{ik}$ is the mixing matrix, which is found from the requirement
of the finiteness of the Green functions of the operator
$[\Psi_i (\phi)]_R$.
Matrix of the anomalous dimensions
$\gamma_{\Psi}^{ik}$ connects with the mixing matrix (in the MS-scheme)
by formula:
\begin{equation}
\label{788}
\gamma_{\Psi}^{ik}(g) =
n_k \gamma_{\phi}(g)\delta^{ik} +
[ 2 \epsilon g \partial_g - M \partial_M]
\frac{Q^{ik}_{1\Psi}}{\epsilon}
\ , \ \mbox{{\rm where}}
\end{equation}
$$
Q_{ik} =
1 + \frac{Q^{ik}_{1\Psi}}{\epsilon} + ...,
$$
$n_k$ - number of fields $\phi$ in the operator $\Psi_k(\phi)$ and
$\gamma_{\phi}(g)$ -- the  anomalous  dimension of field.

Our immediate goal now -- the calculation of the mixing matrix $Q_{ik}$.
Since our consideration restricted by first order of perturbation theory
it is necessary to calculate only two diagrams, shown on Fig.2
(where the black dot denotes the  operator insertion and the external
lines - $\phi$-arguments).
It is convenient to do all calculation in the terms of the generating
function for the composite operator. To avoid a misunderstanding,
it should be stressed that all equalities with the generating
functions  $\Phi(x;a_1,\ldots,a_n)$  is understood in the sense of the
equalities of the coefficient of the expansion in the power series on
$a_i$ of the right and left sides.
   \vspace{0.5cm}

\unitlength=1mm
\special{em:linewidth 0.4pt}
\linethickness{0.4pt}
\begin{picture}(103.00,51.00)
\put(28.00,35.00){\circle{14.00}}
\put(15.00,35.00){$a_s$}
\put(38.00,35.00){$a_m$}
\put(81.00,35.00){$a_s$}
\put(104.00,35.00){$a_p$}
\put(94.50,35.00){$a_m$}
\put(94.00,35.00){\circle{14.00}}
\put(94.00,28.00){\circle*{4.00}}
\put(28.00,28.00){\circle*{5.20}}
\put(94.00,28.00){\circle*{5.20}}
\put(94.00,28.00){\line(0,1){14.00}}
\put(94.00,28.00){\line(-1,-1){9.00}}
\put(94.00,28.00){\line(0,-1){9.00}}
\put(94.00,28.00){\line(1,-1){9.00}}
\put(94.00,42.00){\circle*{2.00}}
\put(94.00,42.00){\line(0,1){9.00}}
\put(94.00,28.00){\line(2,-5){3.67}}
\put(94.00,28.00){\line(-2,-5){3.67}}
\put(28.00,28.00){\line(-1,-1){9.00}}
\put(28.00,28.00){\line(1,-1){9.00}}
\put(28.00,28.00){\line(0,-1){9.00}}
\put(28.00,28.00){\line(2,-5){3.67}}
\put(28.00,28.00){\line(-2,-5){3.67}}
\put(28.00,42.00){\circle*{2.00}}
\put(28.00,42.00){\line(1,1){9.00}}
\put(28.00,42.00){\line(-1,1){9.00}}
\put(28.00,9.00){\makebox(0,0)[cc]{( a )}}
\put(94.00,9.00){\makebox(0,0)[cc]{( b )}}
\put(60.00,5.00){\makebox(0,0)[cc]{Fig.2}}
\end{picture}

Let us calculate the diagram (a).
It has the
following expression in the momenta representation
 ($p = p_s + p_m$) :
\begin{eqnarray}
\lefteqn{ \int \frac{{\rm d}^{D}k}{{\pi}^{D/2}}
\  \frac{\exp{[i v k a_s + i v (p-k) a_m]}}
{k^{2}(p-k)^{2}}=} \nonumber \\
& &=
\int_{0}^{1} {\rm d}s (1 - s)^{D/2 - 2}
\exp{[ivp(sa_s +(1-s) a_m)]}\cdot \nonumber \\
\lefteqn{\cdot  \int_{0}^{\infty} {\rm d}t
t^{1 - D/2}
\exp{[- p^2 t s - v^2(a_s - a_m)^2\frac{1 - s}{4t}]}}
\label{432}
\end{eqnarray}
We need to extract the $\epsilon$ pole contributions ($D=4-2\epsilon$)
to the coefficients of the expansion of this integral in
sources $a_i$.
In the final formula this expression will be stand under sign
of the projector and due to evident property:
$$
\left. P(u , \frac{\partial}{\partial v})
v^2 f(v) \right|_{v = 0} = 0,
$$
it is possible to drop all terms proportional to $v^2$.
Thus for singular part of the diagram (a)
one obtains :
\begin{equation}
\frac{1}
{\epsilon}
\int_0^1 {\rm d}s \>
\exp{[i u p (s a_s + (1-s)a_p)]}. \label{22a}
\end{equation}

The singular part of the diagram (b) is calculated in the similar way.
Finally, we obtain counterterms generated by the first and second diagrams
respectively:
\begin{equation} -\frac{1}{2\epsilon}
\frac{g}{(4\pi)^2} \sum_{i<k} \int_0^1 {\rm d}s \Phi(x;a_1,...,s
a_i + (1-s)a_k,...,s a_i + (1-s)a_k,...,a_n)
\end{equation}
\begin{equation}
\label{6561}
-\frac{1}{2\epsilon} \frac{g}{(4\pi)^4}
\sum_{i<k<m} \int_0^1 {\rm d}s\ {\rm d}t\ t
\Phi(x;a_1,...,{\hat a}_i ,...,{\hat a}_k,...,{\hat a}_m,...,a_n)
\end{equation}
$$
\partial^2 \phi(x + s t a_i + (1-s) t a_k + (1-t) a_m).
$$
where
$
\Phi(x;a_1,...,{\hat a}_i ,...,
{\hat a}_k,...,{\hat a}_m,...,a_n)
$
means that we exclude from $\Phi(x;a_1,\ldots,a_n)$
the fields with "source" ${\hat a}$.
The second diagram generates new
type of operators
\begin{equation}
\label{2222}
\phi(x + a_1 u)\phi(x + a_2 u)...\partial^2 \phi(x + a_{n-2} u).
\end{equation}

Thus, one can see, that the class of the operators defined in Sec.1
is unclosed under renormalization and must be enlarged by the
inclusion of the operators with all possible
$\partial^2\phi$-insertions.
Let us consider the counterterms for the operator with one
$\partial^2\phi$-insertion. It is possible two variants:
\begin{enumerate}
\item  $\partial^2\phi$-insertion get inside a diagram:
in this case the operator $\partial^2$  "erase" a line in a
loop, and we get the zero contribution,
\item  $\partial^2\phi$-insertion is external with respect to
the diagram: in such case the answer is the same as for the operator
without $\partial^2\phi$-insertion.
\end{enumerate}

It is easy to understand, that
the counterterms for the operators with
k $\partial^2\phi$-insertions consist of those of with
k and (k+1) $\partial^2\phi$ - insertions.
This means that mixing matrix $Q_{ik}$ for the set of the
operators in question has "block triangular" form,
block on diagonal being induced by (a)--diagram, and
off--diagonal ones --- by (b)--diagram. It should be noted that
this property is disappeared in the next order of the
perturbation theory.

It is evident that the eigenvalue problem for $Q_{ik}$
is reduced to the eigenvalue problem for
the "main diagonal block".
The renormalization of the composite operator with
coefficient function $\psi$ is given by formula:
$$ [\Psi(x,u)]_R = \left.
\psi(\frac{\partial}{\partial a_1},...,
\frac{\partial}{\partial a_n})
[\Phi(x;a_1,...,a_n)]_R \right|_{a_m = 0}.
$$
where
\begin{equation}
\label{767}
[\Phi(x;a_1,...,a_n)]_R  =  \Phi(x;a_1,...,a_n) +
\end{equation}
$$
+ \frac{1}{2\epsilon} \frac{g}{(4\pi)^2}
\sum_{i<k} \int_0^1 {\rm d}s
\Phi(x;a_1,...,s a_i + (1-s)a_k,...,s a_i +
(1-s)a_k,...,a_n)+\ldots,  $$
where dots denote the counterterms type of (\ref{6561}).

In the result we receive the following eigenvalue problem:
$$
\sum_{i<k} \int_0^1 {\rm d}s
\left.
\psi(\frac{\partial}{\partial a_1},...,
\frac{\partial}{\partial a_n})
\Phi(x;a_1,...,s a_i + (1-s)a_k,...,s a_i + (1-s)a_k,...,a_n)
\right|_{a_m = 0} =
$$
\begin{equation}
= \lambda \left.
\psi(\frac{\partial}{\partial a_1},...,
\frac{\partial}{\partial a_n})
\Phi(x;a_1,...,a_n)\right|_{a_m = 0}.
\end{equation}
By using of the Fourier transformation
$\phi(x)=\int {\rm d}q {\rm e}^{i qx}\phi(q)$
we obtain
\begin{equation}
\sum_{i<k} \int_0^1 {\rm d}s
\psi( z_1,...,
s (z_i + z_k),...,
(1 - s) (z_i + z_k),...,z_n) =
\lambda
\psi(z_1,...,z_n).  \label{psi}
\end{equation}

The corresponding composite operator $\Psi(x;u)$
has in virtue of~(\ref{788}) the follow
anomalous dimension:
\begin{equation}
\gamma_{\Psi} =
\frac{2\lambda}{3} \epsilon + O(\epsilon^2).
\end{equation}

It should be noted  that eq.~(\ref{psi}) defines the
eigenvalues of the mixing matrix $Q_{ik}$ only.
The exact eigenvectors contain
  the admixture of the
operators with  $\partial^2\phi$ - insertions defined by
the counterterms~(\ref{6561}) which can be reconstructed
recurrently.

\setcounter{equation}{0}
\section{Solution for the cases
of two and three fields.}
\label{dim23}

We consider the  equation~(\ref{psi}) from previous part
\begin{equation}
\label{333aa}
\sum_{i<k} \int_0^1 \rm{d}\alpha
\ \psi(z_1,...,\alpha(z_i+z_k),...,(1-\alpha)(z_i+z_k),...,z_n) =
\lambda \psi(z_1,...,z_n).
\end{equation}
It is useful to reformulate problem in terms of a function
$\hat{\psi}(z_1,...,z_n)$. To do  this we perform in
equation~(\ref{333aa}) the transformation~(\ref{5a}) from
the Sec.\ref{descr}.
After transformation~(\ref{5a}) one obtains
\begin{eqnarray}
H \hat{\psi}(z_1,...,z_n) &\equiv &
\sum_{i<k} \int_0^1 \rm{d}\alpha
\hat{\psi}(z_1,...,
\alpha z_i + (1-\alpha)z_k,...,
\alpha z_i + (1-\alpha)z_k,...,z_n) \nonumber \\
H \hat{\psi}(z_1,...,z_n) &= &
  \lambda \hat{\psi}(z_1,...,z_n). \label{334}
\end{eqnarray}
For some purpose the equivalent form of (\ref{334}) is more useful
\begin{equation}
\label{335}
\rm{H} \hat{\psi}(z_1,...,z_n) \equiv
\sum_{i<k} \frac{1}{z_i-z_k}\int_{z_k}^{z_i} \rm{d}s
\hat{\psi}(z_1,...,s,...,s,...,z_n) =
\lambda \hat{\psi}(z_1,...,z_n).
\end{equation}
The important properties of the spectrum have been
established in \cite{1,2}.
The spatial symmetry group of $\rm{H}$ is
the $\rm{SL}(2,C)$ \cite{1}.
Let us define the action of the $\rm{SL}(2,C)$-group
on the functions in the following way
$$
\rm{S}(g)\hat{\psi}(z_1,...,z_n) =
\prod_{i=1}^n (c z_i + d)^{-1}
\hat{\psi}(\frac{a z_1 + b}{c z_1 + d},...,
\frac{a z_n + b}{c z_n + d})
$$
where $g\in SL(2,C),\> g=\left (\begin{array}{cc} a & b \\
 c & d \end{array} \right )$ and
$a d - b c = 1$.
It is not hard to check (using the
representation~(\ref{335})) that operators $\rm{H}$ and
$\rm{S}(g)$
commute for all $g$.
The group $\rm{SL}(2)$ has a three generators
$\rm{S}\ ,\ \rm{S_{+}} \ ,\ \rm{S_{-}}$
$$
\rm{S} \equiv \left.
\frac{\partial}{\partial a} \rm{S}(g) \right|_{g=I}  =
\sum_{i=1}^{n} z_i \frac{\partial}{\partial z_i}
$$
\begin{equation} \label{comm}
\rm{S_{-}} \equiv \left. \frac{\partial}{\partial b}
\rm{S}(g) \right|_{g=I} =
\sum_{i=1}^n \frac{\partial}{\partial z_i} \ , \
\rm{S_{+}} \equiv \left. \frac{\partial}{\partial c}
\rm{S}(g) \right|_{g=I} =
-\sum_{i=1}^n(z_i^2 \frac{\partial}{\partial z_i} + z_i).
\end{equation}
All these generators commute with $\rm{H}$ and have the
following commutation relations
\begin{equation}
[\rm{S_{-}},\rm{S}] = - \rm{S_{-}} \ ,\
[\rm{S_{+}},\rm{S}] = + \rm{S_{+}} \ ,\
[\rm{S_{+}},\rm{S_{-}}] = 2\rm{S} + 1 .
\end{equation}
The operator $\rm{H}$  has the same set of
eigenvectors as $\rm{S}$ (as a commuting operators)
$$
\rm{S}\hat{\psi} = l \hat{\psi}
\Longrightarrow
\rm{H}\hat{\psi} = \lambda \hat{\psi}
$$
The function $\hat{\psi}$ satisfies the equation
$$
\sum_{i=1}^{n} z_i \frac{\partial}{\partial z_i}
\hat{\psi}(z_1,...,z_n) =
l \hat{\psi}(z_1,...,z_n),
$$
and, consequently, $\hat{\psi}$ is the gomogeneous
polynomial of the degree l.
{}From the equalities
$$
\rm{S}\rm{S_{-}}\hat{\psi} =
(l-1)\rm{S_{-}} \hat{\psi} \ ,\
\rm{H}\rm{S_{-}}\hat{\psi} =
\lambda\rm{S_{-}} \hat{\psi}  ,
$$
$$
\rm{S}\rm{S_{+}}\hat{\psi} =
(l+1)\rm{S_{+}} \hat{\psi} \ ,\
\rm{H}\rm{S_{+}}\hat{\psi} =
\lambda\rm{S_{+}} \hat{\psi}
$$
it is seen that operators
$\rm{S_{+}}$ and $\rm{S_{-}}$
are the standard rising and lowering operators.
Then the subspace of the eigenvectors of $\rm{H}$ with some
eigenvalue $\lambda$ is the $\rm{SL}(2)$ - module
generated by highest weight vector $\hat{\psi}$ (vacuum vector),i.e.
vector space spanned by linear combinations of
monomials in the $\rm{S_{+}}$ applied to $\hat{\psi}$.
The highest weight vector $\hat{\psi}$ is defined by
the equation $\rm{S_{-}}\hat{\psi} = 0$ or in
the coordinate form
$$
\sum_{i=1}^n \frac{\partial}{\partial z_i}
\hat{\psi}(z_1,...,z_n) = 0.
$$
This equation is the same equation~(\ref{9a}) from
first part and we obtain the following correspondence:
the highest weight vector $\hat{\psi}$  represents
the canonically conformal composite operator
and all other vectors from $\rm{SL}(2)$ - module
represent a derivatives of this operator.
We obtain that all composite operators with the same
anomalous dimension are divided into two sets.
The operators from first set are the canonically conformal ones
and those from second set
are a derivatives of a conformal
operators.
In this sense the spectrum of anomalous dimensions
is generated by the conformal  operators.

Now we have the eigenvalues problem~(\ref{334})
for the functions $\hat{\psi}(z_1,...,z_n)$
satisfying the following restrictions:
\begin{enumerate}
\item $\hat{\psi}(z_1,...,z_n)$ is the
total symmetric function,
\item $\hat{\psi}(z_1,...,z_n)$ is the
gomogeneous polynomial of degree l,
\item $\hat{\psi}(z_1,...,z_n)$ is the
translation invariant function.
\end{enumerate}
The vector space
$C(n,l)$ of symmetric , gomogeneous
and translation invariant polynomials of degree l
have been considered in Sect.\ref{descr}.

The case $n = 2$ is trivial.
The generating function $\rm{D}(2,x)$ for
the $\rm{dim} C(2,l)$ has
the form~(\ref{77}):
$$
\rm{D}(2,x) =
\frac{1}{1-x^2}
$$
and we obtain that  $C(2,l)$ have nonzero
dimension only for even l.
For even l $\rm{dim}C(2,l) = 1$ and the basis
polynomial is
$$
\hat{\psi}(z_1,z_2) = (z_1 - z_2)^l.
$$
It is evident that this eigenfunction correspond
to the zero eigenvalue.

The case $n=3$ is more complicate.
We consider this case in detail because the same
methods we will be used in the case $n=4$.

All functions which vanishes if any two arguments
coincide correspond of the eigenvalue
$\lambda = 0$ and have
the follow general form
\begin{equation}
\label{88}
\hat{\psi}_0(z_1,\ldots,z_n) =\prod_{i<k}
(z_i-z_k)^2 \phi(z_1,\ldots,z_n),
\end{equation}
where $\phi(z_1,\ldots,z_n)$ symmetric,
translation invariant polynomial
degree ($l - 2\cdot C^n_2$).
Let us the vector space of such functions
$\hat{\psi}_0(z_1,\ldots,z_n)$ denote $C_0(n,l)$.
The generating function  $\rm{D}_0(n,x)$ for
$\mbox{{\rm dim}} C_0(n,l)$ has the form:
\begin{equation}
\label{dim0}
\rm{D}_0(n,x) =
\frac{x^{2\cdot C^n_2}}{\prod_{i=2}^{n}(1-x^i)}.
\end{equation}
It is evident that the difference
$
\rm{D}(n,x) - \rm{D}_0(n,x)
$
(see~(\ref{77}))
is the generating function for the dimensions of the
spaces of the polynomials, which
do not vanish if any two arguments
coincide.

In the case $n=3$
$$
\rm{D}(3,x) - \rm{D}_0(3,x) =
 1+x^2+x^3+x^4+...
$$
Thus we obtain that in vector space $C(3,l)$ exist only
one eigenvector which does not vanish if any two
arguments coincide.

Let us put in the equation~(\ref{335})
for $n=3$ $z_1 = z_2 = z$ then we obtain
the equation on the function of the two variables
$\hat{\psi}(z,z,z_3)$
$$
\frac{2}{z - z_3}\int_{z_3}^{z} \rm{d}s
\ \hat{\psi}(s,s,z) =
(\lambda - 1) \hat{\psi}(z,z,z_3).
$$
The function $\hat{\psi}(z,z,z_3)$ has the
simple form:
$$
\hat{\psi}(z,z,z_3) =
\hat{\psi}(z - z_3,z - z_3,0) =
(z - z_3)^l \hat{\psi}(1,1,0)
$$
(the constant $\hat{\psi}(1,1,0)$ is not equal
to zero because the
function $\hat{\psi}(z,z,z_3) \not\equiv 0$).
If we substitute this function in integral equation
we obtain the eigenvalue
\begin{equation}
\lambda = 1 + \frac{2(-1)^l}{l+1}
\end{equation}
The "true" eigenfunction $\hat{\psi}(z_1,z_2,z_3)$
can be reconstructed
with the help of equation~(\ref{335}).

\setcounter{equation}{0}
\section{Solution for the
case of four fields.}
\label{four}

Let us consider the case $n=4$.
At first, as in the case $n=3$, we calculate
the dimension of the space of polynomials
not vanishing at coinciding of any two arguments.
According to (\ref{77}),(\ref{dim0})
$$
\rm{D}(4,x) - \rm{D}_0(4,x) =
 1+x+x^3+\sum_{L=0}^{\infty} L x^{2L} +
\sum_{L=0}^{\infty} (L-1) x^{2L+1}
$$
Thus one obtains that in vector space $C(4,l)$ there exist
$L$ eigenvectors for $l=2L$ which not vanishes if any two
arguments coincide and $L-1$ such
eigenvectors for $l = 2L+1$ $(L\geq 1)$.

Let us put in equation~(\ref{335}) for $n=4$
$z_1 = z_2 $ and introduce the function of
one variable
$\psi(z)$:
\begin{equation}
\psi(z) \equiv \hat{\psi}(0,0,1,z).
\end{equation}
Using the symmetry properties of $\hat \psi$ one can obtain
$$
\hat{\psi}(z_1,z_1,z_3,z_4) =
 (z_3 - z_1)^l \psi(\frac{z_4 - z_1}{z_3 - z_1}).
$$
Then equation (\ref{335}) takes the following form:
\begin{equation}
\label{555}
\frac{1}{z}\int_{0}^{z} \rm{d}s
\ s^l \psi(\frac{s - 1}{s}) +
\int_{0}^{1} \rm{d}s
\ s^l \psi(\frac{s - z}{s}) +
\frac{\psi(1)(-1)^l}{2(l + 1)}
\frac{1 - z^{l+1}}{1 - z} =
\frac{\lambda - 1}{2}(-1)^l\psi(z),
\end{equation}
where $\psi(1)=0$ for odd $l$.
The function $\psi(z)$ inherits the some
properties of $\hat{\psi}$:
\begin{enumerate}
\item $\psi(z)$ is the polynomial of degree l.
\item $\psi(z)=z^l \psi(\frac{1}{z})$.
\end{enumerate}

It is useful to rewrite the equation~(\ref{555}) as the
functionally-differential equation. For this purpose we introduce
the new function $F(z)$:
$$
F(z) = \int_{0}^{z} {\rm d}s
\ s^l \psi(\frac{s - 1}{s})
$$
Immediately from definition we obtain
$$
F^{\prime}(z) = z^l \psi(\frac{z - 1}{z})
\Longleftrightarrow
\psi(z) = (1 - z)^l F^{\prime} (\frac{1}{1 - z})
\ ;\ F(0) = 0.
$$
In virtue of the property
$\psi(z) = z^l \psi(\frac{1}{z})$
the function $F(z)$ also have the nice symmetry
property
\begin{equation}
F(z) = (-1)^{l+1}F(1-z) + (-1)^lF(1).
\label{prop}
\end{equation}
It is easy obtain from
eq.~(\ref{555}) that
$$
F(1) = (-1)^l \frac{\lambda - 2}{4} \psi(1).
$$
In terms of function $F(z)$  eq.~(\ref{555}) takes the form:
\begin{equation}
\label{556}
\frac{1}{z}F(z) +
z^{l+1} F(\frac{1}{z}) +
\frac{\psi(1)(-1)^l}{2(l + 1)}
\frac{1 - z^{l+1}}{1 - z} =
\frac{\lambda - 1}{2}(-1)^l
(1 - z)^l F^{\prime} (\frac{1}{1 - z}).
\end{equation}
where
$F(z)$ is the polynomial of degree
(l+1) and $F(0) = 0$.

\subsection{The eigenfunctions with
eigenvalues $\lambda = 1$.}
\label{zero}

Let us put $\lambda = 1$.
Then we obtain the functional equation:
\begin{equation}
\label{z1}
\frac{1}{z}F(z) +
z^{l+1} F(\frac{1}{z}) +
\frac{\psi(1)(-1)^l}{2(l + 1)}
\frac{1 - z^{l+1}}{1 - z} = 0.
\end{equation}
The solutions of this equation differs for
the cases odd and even $l$ but they have
one common property: for all these
solutions  $\psi(1) = 0$.
Let us  consider at first the case of odd $l$
($l = 2L + 1$)
\begin{equation}
\label{333}
F(z) + z^{2L+3} F(\frac{1}{z}) = 0.
\end{equation}
It is easy to check by direct calculation that functions
\begin{equation}
\label{222}
F_{a,b}(z) = [z(1 - z)]^a[1 - z(1 - z)]^b
\end{equation}
are the solutions of (\ref{333}) if the following
conditions are fulfilled
$$
a \ {\rm is\  odd\  and} \ 3a + 2b = 2L +3 \ ,\ a + b \leq L.
$$
The parameters $a$ and $b$ take the following values:
$$
L = 3M \ , \ a = 2k +1 \ ,
\ b = 3 (M - k) \ \rm{where} \ k = 1,...,M ;
$$
\begin{equation}
\label{567}
L = 3M + 1 \ , \ a = 2k +1 \ ,
\ b = 3 (M - k) + 1 \ {\rm where} \ k = 1,...,M ;
\end{equation}
$$
L = 3M + 2 \ , \ a = 2k +1 \ ,
\ b = 3 (M - k) - 1 \ {\rm where} \ k = 0,...,M - 1 ,
$$
therefore the dimension of the eigenspace with $\lambda = 1$
is equal to M for $L = 3M + r \ (r = 0 , 1 , 2)$.

In the case of the even l ($l = 2L$) the equation
\begin{equation}
\label{333ab}
F(z) + z^{2L+2} F(\frac{1}{z}) = 0
\end{equation}
with the help of the ansatz
$$
F(z) = (1-2z)(-2 - z(1-z)) G(z) =
(1-2z)(z-2)(z+1) G(z)
$$
(where $G(z) = G(1-z)$)
reduces to the one for odd case
$$
G(z) + z^{2L-1} G(\frac{1}{z}) = 0.
$$
The solutions are:
\begin{equation}
\label{222a}
G_{a,b}(z) = [z(1 - z)]^a[1 - z(1 - z)]^b
\end{equation}
where
$$
a \ {\rm is\  odd\  and} \ 3a + 2b = 2L -1 \ ,\ a + b \leq L-2.
$$
(from the condition $\psi(1) = 0$
follows that $F(z)$ is polynomial
of the degree $\leq 2L - 1$ and therefore $a + b \leq L-2$)
and
$$
 a = 2k +1 \ ,
\ b = 3 (M - k) - 2 \ {\rm where} \ k = 1,...,M - 1 \ {\rm for} \ l=3M
$$
$$
  a = 2k +1 \ ,
\ b = 3 (M - k) - 1 \ {\rm where} \ k = 1,...,M - 1 \ {\rm for} \ L = 3M + 1
$$
$$
a = 2k +1 \ ,
\ b = 3 (M - k) \ {\rm where} \ k = 1,...,M \ {\rm for} \ L = 3M + 2.
$$
The dimension of the eigenspace with $\lambda = 1$
is equal to M-1 for $L = 3M + r \ (r = -1 , 0 , 1 )$ and $M > 1$.

\subsection{The solutions for the
nondegenerate eigenvalues.}

In this Section we limit ourselves to the determination
of the nondegenerate
eigenvalues ($\lambda \neq 1$).
In this case we obtain the interesting
result - the equation~(\ref{556})
on the function $F(z)$ with symmetry property~(\ref{prop})
 for $\lambda \neq 1$
is equivalent to the linear differential equation
of the third order
$$
 - (z(1-z))^2 \frac{{\rm d}^3 F(z)}{{\rm d}z^3} +
(l - 1)z(1-z)(1-2z) \frac{{\rm d}^2 F(z)}{{\rm d}z^2} +
$$
\begin{equation}
\label{9876}
+ [z(1-z)(l(l-1) - \mu^2 - \mu) + l+1+\mu^2 ]
\frac{{\rm d}F(z)}{{\rm d}z} -
\end{equation}
$$
- \mu^2 (l+2)\frac{1-2z}{z(1-z)} F(z) =  P(z)
$$
where
$\mu \equiv \frac{2}{\lambda - 1}$ and
$$
P(z) = [(-1)^l(1-z)^{l+3} + z^{l+3}]
\mu(\mu - 1)(\mu + 2)\frac{\psi(1)}{2(l+1)} +
$$
$$
+[(-1)^l(1-z)^{l+2} + z^{l+2}]
(F(1)\mu^2(\mu + 1) +
 \frac{\psi(1)}{2(l+1)}
\mu(l+3 - \mu - \mu^2)) +
$$
$$
+ [(-1)^l z \frac{1-z^{l+1}}{1-z} +
 (1-z) \frac{1-(1-z)^{l+1}}{z}]\mu^3 \frac{\psi(1)}{2(l+1)}+
$$
$$
+ F(1)\mu^2(\mu+1) +
 \mu^3 \frac{\psi(1)}{2}[1 + (-1)^l] +
\mu^2 (l+2)F(1) \frac{z}{1-z}.
$$
The derivation of this equation
we give in Appendix B.

In the case of odd $l$ $\psi(1)=F(1)=0$,
and consequently $P(z)=0$.
The corresponding gomogeneous
equation
has the respectively simple
solutions.
For generality we shall consider
case of the  arbitrary l (real or complex).
This equation belongs to the Fuchs class
and have three singular points
$z = 0,\ z = 1,\ z = \infty$.
In the neighbourhood of the singular point $z_0 \neq \infty$
the solution of this equation has the
simple form
$$
F(z) = (z-z_0)^\alpha {\hat F}(z),
$$
where the function ${\hat F}(z)$ is regular
in the point $z_0$, i.e.
${\hat F}(z) = \sum_{n=0}^{\infty}{\hat F}_n (z - z_0)^n$.
The expansion of $F(z)$ in the vicinity of the point $z_0 = \infty$
can be written in the similar way
$$
F(z) = z^\alpha {\hat F}(z),
$$
where ${\hat F}(z) = \sum_{n=0}^{\infty}{\hat F}_n z^{-n}$.
For the point $z_0 = 0$ one obtains the following
characteristic equation for $\alpha$
$$
\alpha^3 - (l+2)\alpha^2 - \mu^2 \alpha + \mu^2 (l+2) = 0
$$
with solutions
$\alpha = l+2 \ ,\ \alpha = \mu \ ,\ \alpha = - \mu $.
For the point $z_0 = \infty$ we have
$$
\alpha(\alpha^2 - (2l+1)\alpha +l(l+1) - \mu - \mu^2) = 0
$$
and therefore
$\alpha = 0 \ ,\ \alpha = l - \mu \ ,\ \alpha = l + 1 + \mu $.
In virtue of the evident symmetry
($z \longrightarrow 1 - z$) of the equation~(\ref{9876})
the characteristic equation
for the point $z_0 = 1$ is the same as for point $z_0 = 0$.
Let us  look for  the
solutions  in the form
$F(z) = f(z(1-z))$.
The eq.~(\ref{9876}) with $P(z)=0$ in terms of
the variable $t = z(1-z)$ is rewritten as
$$
t^2(1-4t) \frac{{\rm d}^3 f(t)}{{\rm d}t^3} -
t(l - 1 + t(10-4l))\frac{{\rm d}^2 f(t)}{{\rm d}t^2} +
$$
\begin{equation}
\label{9878}
+ [t(2(l-1) - l(l-1) + \mu^2 + \mu) - l - 1 - \mu^2 ]
\frac{{\rm d}f(t)}{{\rm d}t} +
\frac{\mu^2 (l+2)}{t} f(t) =  0.
\end{equation}
The solutions of this equation have the form
$$
f(t) = t^{\alpha} \sum_{n = 0}^{\infty} f_n t^n.
$$
Eq.(\ref{9878}) leads to the simple recurrent relations for the
coefficients $f_n$:
\begin{equation}
\label{9879}
f_{n+1} =
\frac{(n+\alpha)(2n+2\alpha-l+\mu)(2n+2\alpha-l-1-\mu)}
{(n+\alpha-l-1)(n+\alpha+1-\mu)(n+\alpha+1+\mu)} f_n.
\end{equation}
Substituting the corresponding values for $\alpha$
we obtain three solutions:
\begin{equation}
\label{9880}
\alpha = \mu \ ,\  f_{n+1} =
\frac{(n+\mu)(n+(3\mu-l)/2)(n+(\mu-l-1)/2)}
{(n+2\mu+1)(n+\mu-l-1)} \frac{4}{n+1}f_n.
\end{equation}
\begin{equation}
\label{9881}
\alpha = - \mu \ ,\  f_{n+1} =
\frac{(n-\mu)(n-(3\mu+l+1)/2)(n-(\mu+l)/2)}
{(n-2\mu+1)(n-\mu-l-1)} \frac{4}{n+1}f_n.
\end{equation}
\begin{equation}
\label{9882}
\alpha = l+2 \ ,\  f_{n+1} =
\frac{(n+l+2)(n+(\mu+l+4)/2)(n+(l+3-\mu)/2)}
{(n+l-\mu+3)(n+l+\mu+3)} \frac{4}{n+1}f_n.
\end{equation}
These solutions (with normalization $f_0 = 1$)
can be represent in terms of
generalized hypergeometric function:
$$
F(a,b,c;d,e;x) = \sum_{n=0}^{\infty}
\frac{(a)_n(b)_n(c)_n}{(d)_n(e)_n}\frac{x^n}{n!}
$$
where
$$
(a)_0 = 1\ , \  (a)_n = a(a+1)...(a+n-1) =
\frac{\Gamma(a+n)}{\Gamma(a)}
$$
in the following way
 \begin{equation} \label{9883}
 f_{\mu}(t) =
t^{\mu}F(\mu,(3\mu-l)/2,(\mu-l-1)/2;2\mu+1,\mu-l-1;4t),
\end{equation}
\begin{equation}
\label{9884}
f_{-\mu}(t) = t^{-\mu}
F(-\mu,-(3\mu+l+1)/2,-(\mu+l)/2;-2\mu+1,-(\mu+l+1);4t),
\end{equation}
\begin{equation}
\label{9885}
f_{l+2}(t) = t^{l+2}
F(l+2,(\mu+l+4)/2,(l+3-\mu)/2;l+3-\mu,l+3+\mu;4t),
\end{equation}

First we consider the case of the odd l ($l = 2L + 1$).
In this case we can obtain the exact solution of the problem
and moreover this exact solution give  us the key
for the understanding of the structure of the spectrum
for all l.

The eigenvalue $\mu$ is defined by the following
requirement: for given $\mu$ the polynomial solution
must exist.
(The requirement  $\psi(1) = 0$ decrease the degree of the
polynomial $f(t)$ by one ($f(t) = \sum_{n=1}^{L} f_n t^n$)).

The requirement of the polynomiality immediately leads to
the following set of eigenvalues:
$$
L = 3M, \  \ \mu = 2k +1, \  \mbox{{\rm where}} \ k = 0,...,M-1
$$
$$
\mbox{{\rm and}} \  \mu = -2k, \  \mbox{{\rm where}} \ k = 1,...,M ;
$$
$$
L = 3M +1, \  \ \mu = 2k +1, \  \mbox{{\rm where}} \ k = 0,...,M
$$
$$
\mbox{{\rm and}} \  \mu = -2k, \ \mbox{{\rm where}} \ k = 1,...,M ;
$$
$$
L = 3M +2, \   \ \mu = 2k +1, \  \mbox{{\rm where}} \ k = 0,...,M
$$
$$
\mbox{{\rm and}} \  \mu = -2k, \  \mbox{{\rm where}} \ k = 1,...,M + 1 .
$$
and therefore exist $2M + r$ nondegenerate eigenvalues
for $L = 3M + r \ (r = 0 , 1 ,2)$.

Thus for $L=3M+r$ ($r=0,1,2$) there exists $L$ solutions of
eq.(\ref{555}) ($M$ solutions with $\lambda = 1$,  $2M+r$ ones
with $\lambda = 1 + \frac{2}{k}(-1)^{k+1}$,
$k = 1,2,...,2M + r$).
On the other hand, the dimension
of the vector space of the
gomogeneous, symmetric and translation
invariant polynomials $C(4,2L+1)$,
which does not vanish if any two
arguments coincide, is equal to $L-1$.
Hence, going from eq.~(\ref{335}) to eq.~(\ref{555}) we got one
additional solution. This fact has simple explanation. There
are functions $\psi(z)$ which cannot be obtained from anyone
$\hat \psi(z_1,..z_4)\in C(4,2L+1)$ by procedure described in
Sec.\ref{four}. It is not hard to understand which solution must
be deleted from spectrum.
As seen from
eq.~(\ref{335}) it is always possible to reconstruct
$\hat \psi(z_1...z_4)$ corresponding to solution of eq.~(\ref{555})
with nonzero $\lambda$. So, the unnecessary solution
of eq.~(\ref{555}) corresponds to $\lambda = 0$ ($\mu = 2$).

Unlike the case $l=2L+1$ we are failed to obtain the analytical
expression for the spectrum when $l=2L$. However, the numerical
calculations reveal some interesting properties of spectrum.

In the corse of calculation we find the following facts.
First of all, as and for  odd case there exists solution of
eq.(\ref{9876}) with $\mu=2, \>\> (\lambda =0)$ (it is the exact
result for both cases) which does not correspond to any
solution of eq.(\ref{335}) and consequently must be deleted from
the spectrum.

Further, let us pick out the following intervals on the real
axis: $ I^+_i=[2i,2i+1]$,   $I^-_i=[-2i,-2i+1],\>\>i=2,3,\ldots$
and $I_1=[1,2],\>\> I_2=[1,2],\>\> I_3=[3,4]$. Then for any
given $M\geq 1, L\geq 4$, ($L=3M+r, \ \ \ r=0,1,2$) there are
three eigenvalues $\mu_1(L),\ \mu_2(L), \mu_3(L)$ which lay in
the intervals $I_1, \ I_2$ and $I_3$ correspondingly.
The others nonzero eigenvalues are situated by the following
way: \\
$
\mu^+_i(L) \in I^+_i,$, \\
where
$i=2,\ldots, M+1,$
for $L=3M+1, \ \ L=3M+2$ \\ and
$  i=2,\ldots, M+2,$
for $L=3M$ \\
$
\mu^-_i(L) \in I^-_i,$, \\
where
$i=2,\ldots, M+2,$
for $L=3M+1, \ \ L=3M+2, \ \L=3(M+1)$  \\
The set of zero eigenvalues is described in Sec.\ref{zero}

In the course of a numerical calculations we have observed the
following properties of eigenvalues :
\begin{enumerate}
\item For any given $i$ at $L\rightarrow\infty$ we have
$$
\mu^+_i(L)\mathop{\longrightarrow}_{L\rightarrow\infty}  2i+1;  \ \
\mu^-_i(L)\mathop{\longrightarrow}_{L\rightarrow\infty}  -2i  \ \
$$
and
$$
\mu_1(L)\mathop{\longrightarrow}_{L\rightarrow\infty}  1  \ \
\mu_2(L)\mathop{\longrightarrow}_{L\rightarrow\infty}  2  \ \
\mu_3(L)\mathop{\longrightarrow}_{L\rightarrow\infty}  3  \ \
$$
\item The sequences $\{\mu^+_i(m)\}, \ \  \{\mu^-_i(m)\}, \ \
\{\mu_1(m)\}, \ \
\{\mu_3(m)\}$ are monoton.
$$
\mu^+_i(m) < \mu^+_i(m+n), \ \ \mu_1(m) < \mu_1(m+n), \ \ n>0,
$$
$$
\mu^-_i(m) > \mu^+_i(m+n), \ \ \mu_3(m) < \mu_3(m+n), \ \ n>0,
$$
Beginning with $L=7$ $\{\mu_2(m)\}$ tends to its limit
monotonically also.
\item At the increasing $L$ a new eigenvalues appear close
enough to the ends of intervals $I^\pm_i$, to left one for
positive $\mu$, and to right one for negative.
In other words
$$
[\mu^+_{M-i}(L)-2(m-i)]
\mathop{\longrightarrow}_{L\rightarrow\infty}  0;
$$
$$
[\mu^-_{M-i}(L)+2(m-i)-1]
\mathop{\longrightarrow}_{L\rightarrow\infty}  0;
$$
here $i$-fixed, and $L=3M+r$.
\end{enumerate}

So, we have the following picture, in any interval
$I_i^+(I_i^-)$ there exists only one eiegenvalue which appears
at some $L>L_0$ at the left (right) boundary of the interval
$I_i^+(I_i^-)$ and then moves, slowly enough, to the opposite
boundary.  (We collect in Table~1 a few first eigenvalues
for different values of $L$ to demonstrate their evolution with $L$).

\setcounter{equation}{0}
\section{The general structure of spectrum}
\label{tree}

In the previous section the spectrum of the anomalous dimension for the
case $n=4$ has been described.
Though we have failed to obtain the analytical expression for all
eigenvalues we have note that in the limit  $l\to\infty$ the
structure of spectrum is strongly simplified. In this section
we consider the structure of the spectrum for nonzero
eigenvalues for arbitrary $n$. All eigenvectors  with $\lambda=0$
were obtained in \cite{2}.

In paper~\cite{2} it has been shown, that
the eigenvalue problem~~(\ref{335}) is equivalent to one
for the hermitean positive definite operator $H$.
\begin{equation}
H =\frac{1}{2} \sum_{n=0}^{\infty}\frac{1}{n+1} H_n^\dagger H_n=
\frac{1}{2} \sum_{n=0}^{\infty}\frac{1}{n+1}
\sum_{i=0}^{n} a_i^\dagger a_{n-i}^\dagger
\sum_{j=0}^{n} a_j a_{n-j}
\label{oper}
\end{equation}
(where $a_i^\dagger, a_i$ -- are creation and annihilation
operators with the standard commutation relations
 $ [a_i,a_k^\dagger]=\delta_{ik}$)
acting in the Fock space.
To the every symmetric polynomial
 $$ \psi(z_1,
\ldots,z_n)=\sum_{\{j_i\}}c_{j_1,\ldots,j_n}z_1^{j_1}\ldots
z_n^{j_n}
$$
the vector from the Fock space corresponds
$$
\psi=\sum_{\{j_i\}}c_{j_1,\ldots,j_n}a^\dagger_{j_1}\ldots
a^\dagger_{j_n} \psi_0,
$$
where $\psi_0$ -- vacuum vector.

The operators $S,S^\pm$~(\ref{comm})  in terms of
$a_i,a_i^\dagger$  are expressed in the following way:
\begin{eqnarray}
S &=& \sum_{j=0}^\infty j\cdot a_j^\dagger\ a_j \nonumber \\
S^- &=& \sum_{j=0}^\infty (j+1)\cdot a_j^\dagger\ a_{j+1} \label{sl2}
\\ S^+ &=& -\sum_{j=0}^\infty (j+1)\cdot a_{j+1}^\dagger\ a_j
\nonumber \end{eqnarray}

We will say that the vector from the Fock space belongs to the
 $n$- level,  if it contains $n$ creation operators.
(Note, that the number of the creation operators is equal to the
number of the variables in the function
$\hat \psi(z_1,\ldots,z_n)$.)
It is evident from eq~(\ref{oper}) that subspace
of the vectors from $n$ level is invariant subspace of the
operator $H$.

Now we prove the following theorem.
\vspace{0.5cm}

{\bf Theorem:} Any eigenvalue of the operator $H$ from the
$n$-level appears on the $(n+i)$-level either  as a accumulation
point of the spectrum or as the exact infinitely degenerate
eigenvalue.

At first, we prove the following simple Lemma.
\vspace{0.5cm}

{\bf Lemma:}
Let for hermitean matrix $A$ there exist vector
$\phi$ ($||\phi||=1$) and number $\lambda$ such that
$$
||(H-\lambda)\phi||\leq \epsilon.
$$
Then at least for one eigenvalue $\lambda_A$ of matrix A
the following inequality
 $|\lambda_A-\lambda|\leq\epsilon$ is fulfilled.
Indeed,
$$
\epsilon\geq||(A-\lambda)\phi||=||\sum_{k}(A-\lambda)c_k\psi_k||=
$$
$$
=(\sum_{k}(\lambda_{k}^A-\lambda)^2 c_k^2)^{1/2}\geq
min|\lambda_{k}^A-\lambda|\cdot(\sum_{k}c_k^2)^{1/2}=
min|\lambda_{k}^A-\lambda|
$$

This Lemma admits the evident generalization.
If for hermitean matrix $A$ there exist set of orthogonal vectors
$\phi_k$ ($||\phi_k||=1$) and number $\lambda$ such that
$$
||(H-\lambda)\phi_k||\leq \epsilon.
$$
Then matrix $A$ has at least $k$ eigenvectors
with eigenvalues  $\lambda_k$ such that
$|\lambda_k-\lambda|\leq(k+1)\epsilon$.

Let vector $\psi_n$ belongs to $n$-level and
\begin{equation}
H\psi_n=\lambda\cdot\psi_n, \ \  \   \
S\psi_n=k\psi_n.
 \label{hpsi}
\end{equation}

Let us transfer the vector $\psi_n$ from $n$ level to $(n+i)$
one in according with the following formula
$f_{n+i}=(a_p^\dagger)^i\psi_n$.
( "kg(a+(, $ ||(H-\lambda)f_{n+i}||.  $
Then taking into account that
$[H_j,a_p^\dagger]=0$ at $p>j$, and $||H^\dagger_j
a_{i_1}^\dagger \ldots a_{i_n}^\dagger||\leq \sqrt{(n+2)!\cdot
j}$ we obtain:
\begin{eqnarray}
\lefteqn{||(H-\lambda)f_{n+i}||=} \nonumber \\
&&=\frac{1}{2}||(\sum_{j=p}^{p+k}\frac{1}{j+1}
H^\dagger_j\cdot H_j +\frac{1}{2p+1}H^\dagger_{2p}H_{2p})
f_{n+1}||\leq \frac{C(k,n,i)}{\sqrt{p}}
||f_{n+i}||,
\label{eqh}
\end{eqnarray}
where  constant $C(k,n,i)$ depends only from $k$,$n$ and $i$.
For the sake of brevity we explain this statement
on the concrete example.The generalization is straightforward.

Let us choose the initial vector in the form:
$(a_0^{\dagger})^n\psi_0$, $H(a_0^\dagger)^n\psi_0=C^n_2\cdot
(a_0^\dagger)^n\psi_0$.
It is evident that only two terms
from $H$, namely
$H_0^\dagger H_0$ and $H_p^\dagger H_p$,
give the nonzero result at acting on the vector
$f_{n+1}$.
Therefore we obtain
\begin{eqnarray}
\lefteqn{ ||(H-C^n_2)a_p^\dagger a_0^{\dagger
n}\psi_0|| =\frac{1}{2(p+1)} ||H^\dagger_p H_p
a_p^\dagger(a_0^\dagger)^n\psi_0||=} \nonumber \\
&=&\frac{n}{(p+1)}
||H^\dagger_p (a_0^\dagger)^{n-1}\psi_0||
\leq
\frac{n}{(p+1)}\sum_{k=0}^p ||a^\dagger_k a^\dagger_{p-k}
(a_0^\dagger)^{n-1}\psi_0||\leq \frac{2n\cdot \sqrt{n!}}{\sqrt{p+1}}
\ \ \  \ \ \ \
\end{eqnarray}

Thus for the any eigenvector
 $\psi_n$ and arbitrary $\epsilon$
 it is possible to choose
 such $p$, that $||(H-\lambda)f_{n+1}||\leq\epsilon$.

Therefore (see Lemma),  we conclude that on $n+1$ level
there exists the eigenvector with eigenvalue $\lambda_n$
such that $|\lambda-\lambda_n|\leq\epsilon$.
Acting on this vector sufficiently enough times by the operator
$S^{-}$ one obtains the conformal vector $\zeta$
($S^{-}\zeta = 0$) with the same eigenvalues
($S\zeta=l\zeta, \ \l\leq p+k$).

Analogously, considering the orthogonal vectors
$a_{p-m}(S^+)^m \psi_n$
($k,m\ll p$) we can prove the existence on the
$n+1$ level m conformal eigenvectors with the
eigenvalues $\lambda_i$,
for which $\lambda$ ($|\lambda- \lambda_i|\leq
C/\sqrt{p}$).  \vspace{0.5cm}

Let us prove  theorem.
Let suppose that  $\lambda$  has the finite degeneracy and is
not the accumulation point of the spectrum on $n+i$ level.
Then there exists the number $\delta$ such that the distance from
$\lambda$ to any other points of spectrum is greater than
$\delta$. On the other hand, using described above procedure
(choosing large enough $p$)  one can state that on the $(n+i)$
level operator $H$ has at least  $M$ eigenvectors with the
eigenvalues $\lambda_i$ for which $|\lambda_i-\lambda|<\epsilon$.
Moreover, $M$ always can be chosen greater than $N$, that
contradicts to the initial assumption.

Let us demonstrate how this theorem can be applied for the
qualitative explanation of the spectrum's structure.
Let us begin  from level $n=2$. In this case only one nonzero
eigenvalues (equal to one) exists.In the according with the
mentioned above on the level $n=3$ there are the sequence of the
eigenvectors with the eigenvalues
$\lambda_l=1+\frac{2(-1)^l}{l+1}$,
tends to  $1$ at $l\to \infty$
(see Sec.~\ref{dim23}).
Analogously, on level
$n=4$
for even $l$ there exist the sequences of the eigenvalues
which converge to $\lambda_k=1+\frac{2(-1)^k}{k+1}$ at $l\to
\infty$, and for odd $l$ the corresponding eigenvalues
$\lambda_k$ have the infinite degeneracy.
Moreover, in the both cases in the spectrum the infinitely
degenerate eigenvalue
$\lambda=1$ from the level $n=2$ is present.

In the result the qualitative structure of spectrum can be
described in the following way.
Every eigenvalue $\lambda_n = C^n_2$ for
eigenvector $(a^\dagger_0)^n\psi_0$  from the level $n$,
($S\psi_n=0$), is a founder of the whole "genealogical tree".
 On the level $n+1$ it generate infinite sequence
of the eigenvalues. At  $l\to \infty$ this sequence tends
to its "ancestor" $\lambda_n$.  On the level
 $n+2$ every eigenvalue from this sequence
generates  own  analogous one, which converges already to its
"ancestor"  at
$l\to \infty$ and so on.
 The results for
levels  $n=2,3,4$ allows us to formulate
the hypothesis that the representative from this sequence
exist for every  $l$.
 It should be stressed that there exist the sequences of
 the eigenvalues which has  not "ancestor" ($\mu_2(k)$ for
 example).

Note, that analogously observations have been made by S.Kehrein
(private communication) on the base of numerical calculation.

Thus, "almost every"  eigenvalue can be coded
by the set of natural numbers describing its position on the
"genealogical tree".

\vspace {0,5cm}
\centerline{\bf Acknowledgments}
\vspace {0,5cm}

Authors are grateful to Prof.~A.N.Vasil`ev and Dr.~Y.M Pis'mak
for stimulating
discussions and critical remarks.
S.Derkachov would like to
thank Dr.A.A.Lobaschov for
stimulating discussion on first stage of the work.
We would like to thank Dr. S.Kehrein for the interest to the
work.

The work was supported by Grant  95-01-00569a of Russian Fond of Fundamental
Research and by Royal Swedish Academy of Sciences
 and is carried under the research program of International Center for
 Fundamental Physics in Moscow.

\setcounter{equation}{0}
\renewcommand{\theequation}{A.\arabic{equation}}
\section{Appendix A.}

In this Appendix we derive the differential equation on the
coefficient function of the canonically conformal
composite operator.
In terms of the composite operators $\Psi(x,u)$
we may rewrite the transformation law (1) in more compact form
\begin{equation}
\delta_{\alpha}\Psi(x,u) = \left.
L_{\alpha}(u,\frac{\partial}{\partial v}) \Psi(x,v)\right|_{v=0}
\end{equation}
In virtue of the locality $L_{\alpha}$, this differential
operator can be written in the form
\begin{equation}
L_{\alpha}(u,s) = i \int {\rm e}^{ipx} \alpha_i(p)
l_i(u,s,p,q){\rm d}p \ ,\
q \equiv -i \frac{\partial}{\partial x} \ ,\
s \equiv \frac{\partial}{\partial v}
\end{equation}

For operators with the transformation law (2)
(for general $\alpha$) we have
\begin{equation}
L_{\alpha}(u,s) = {\rm e}^{us}
(\frac{\Delta_{\Psi}}{D}\partial\alpha +
\alpha\partial + (u\partial)(\alpha s)).
\end{equation}
and
\begin{equation}
\label{1a}
l_i(u,s,p,q) = {\rm e}^{us}
(\frac{\Delta_{\Psi}}{D}p_i + q_i + (up) s_i))
\end{equation}
For the operators $\Psi(x,u)$ it is not difficult
to calculate the $l_i(u,s,p,q)$.
On the one hand
$$
\delta_{\alpha}\Psi(x,u) = \left.\psi(\frac{\partial}{\partial a_1},...,
\frac{\partial}{\partial a_n})
\delta_{\alpha}\Psi(x;a_1,...,a_n)\right|_{a_m=0}
$$
and
$$
\delta_{\alpha}\Psi(x;a_1,...,a_n) =
i \int {\rm d}p {\rm e}^{ipx} \alpha_i(p)
\left. \sum_{k=1}^n {\rm e}^{i a_k up}
[-i \frac{\partial}{\partial x_i^k} + \frac{\Delta}{D}p_i]
\Psi(x;a_1,...,a_n)\right|_{x^k=0}  =
$$
$$
= i \int {\rm d}p {\rm e}^{ipx} \alpha_i(p)
\int \prod_{m=1}^n {\rm d}q^m {\rm e}^{i q^m (x+a_m u)} \phi(q^m)
\sum_{k=1}^n {\rm e}^{ia_k up}
[q_i^k + \frac{\Delta}{D}p_i].
$$
In last equality we have used the Fourier transformation
$\phi(x)=\int {\rm d}q {\rm e}^{i qx}\phi(q)$.
{}From these formulae we obtain
$$
\delta_{\alpha}\Psi(x,u) =
i \int {\rm d}p {\rm e}^{ipx} \alpha_i(p)
\int \prod_{m=1}^n {\rm d}q^m {\rm e}^{i q^m x} \phi(q^m)
\sum_{k=1}^n [q_i^k + \frac{\Delta}{D}p_i]
\psi(i q^1 u,...,i q^k u + i up,...,i q^n u).
$$
On the other hand use the definition of the operator $l_i$
we have
$$
\delta_{\alpha}\Psi(x,u) =
\left. i \int {\rm d}p {\rm e}^{ipx} \alpha_i(p)
l_i(u,\frac{\partial}{\partial v} , p ,-i \frac{\partial}{\partial x})
\int \prod_{m=1}^n {\rm d}q^m {\rm e}^{i q^m x} \phi(q^m)
\psi(i q^1 v,...,i q^n v)\right|_{v=0}.
$$
{}From last two equalities we obtain
\begin{equation}
\label{2a}
\left. l_i(u,\frac{\partial}{\partial v} , p ,\sum q^k)
\psi(i q^1 v,...,i q^n v)\right|_{v=0} =
\sum_{k=1}^n (q_i^k + \frac{\Delta}{D}p_i)
\psi(i q^1 u,...,i q^k u + i up,...,i q^n u)
\end{equation}

This equation allows us to determine the exact form
of the composite operator in according with its
transformation property.
For the canonically conformal  composite
operator $l_i$ must coincide with~(\ref{1a})  for
conformal $\alpha$. From this requirement
the definite restrictions on the $l_i(u,s,p,q)$  follow:
$$
\left. l_i(u,s,p,q)\right|_{p=0} =
{\rm e}^{us}q_i \ ; \
\left. (\frac{\partial l_i(u,s,p,q)}{\partial p_k} -
\frac{\partial l_k(u,s,p,q)}{\partial p_i})
\right|_{p=0} =
{\rm e}^{us}(u_k s_i - u_i s_k)
$$
$$
\left. \frac{\partial l_i(u,s,p,q)}{\partial p_i}
\right|_{p=0} =
{\rm e}^{us}(\Delta_{\Psi} + us) \ ; \
\left. \frac{\partial^2 l_i(u,s,p,q)}
{\partial p_k \partial p_k}
\right|_{p=0} = 2 \left. \frac{\partial^2 l_k(u,s,p,q)}
{\partial p_i \partial p_k}
\right|_{p=0}.
$$
These restrictions on the $l_i$ and the equation~(\ref{2a})
lead to the expression for
the canonical dimension $\Delta_{\Psi} = n \Delta$
and to the equation for the symmetric and
gomogeneous function $\psi(z_1,...,z_n)$
\begin{equation}
\sum_{m=1}^n (z_m \frac{\partial^2}{\partial z_m^2} +
\Delta \frac{\partial}{\partial z_m}) \psi(z_1,...,z_n) = 0.
\end{equation}

\setcounter{equation}{0}
\renewcommand{\theequation}{B.\arabic{equation}}
\section{Appendix B.}

In this Appendix we derive the differential equation~(\ref{9876}).
Let us start from  the equation
\begin{equation}
\label{10000}
\frac{1}{z}F(z) +
z^{l+1} F(\frac{1}{z}) +
\frac{\psi(1)(-1)^l}{2(l + 1)}
\frac{1 - z^{l+1}}{1 - z} =
\frac{1}{\mu}(-1)^l
(1 - z)^l F^{\prime} (\frac{1}{1 - z}).
\end{equation}
The function $F(z)$ has the property
$$
F(z) = (-1)^{l+1}F(1-z) +
(-1)^l F(1).
$$
On the first step we perform the
change of variables
$z \longrightarrow \frac{z-1}{z}$
in the main equation~(\ref{10000}) and obtain
\begin{equation}
\label{10001}
z^{l+1} F(\frac{1}{z}) =
\frac{1}{\mu} (1-z) F^{\prime}(z) +
(-1)^{l+1}\frac{(1-z)^{l+2}}{z} F(\frac{1}{1-z}) +
\end{equation}
$$
+ F(1)z^{l+1} + F(1)(-1)^l\frac{(1-z)^{l+2}}{z} -
\frac{\psi(1)}{2(l+1)}[(1-z)z^{l+1} + (-1)^l (1-z)^{l+2}].
$$
Then,  substituting the expression for
$z^{l+1} F(\frac{1}{z})$ from~(\ref{10001})
in~(\ref{10000}) one obtains
\begin{equation}
\label{10002}
z^{l} F^{\prime}(\frac{1}{z}) +
\mu \frac{z^{l+2}}{1-z}F(\frac{1}{z}) =
z F^{\prime}(z) - \mu \frac{1}{1-z}F(z)+
\mu F(1)[\frac{(-1)^l}{1-z} +
(1-z)^{l+1} + \frac{(-z)^l}{1-z}]+
\end{equation}
$$
+ \mu \frac{\psi(1)}{2(l+1)}
[(-1)^l\frac{1 - (1-z)^{l+1}}{z} -
(-z)^{l+2} - z(1-z)^{l+1}].
$$
On the third step we express
$F(\frac{1}{1-z})$ from the equation~(\ref{10001}) and
substitute it in the equation~(\ref{10000}).
Then the equation similar to one~(\ref{10002}) can be obtained
\begin{equation}
\label{10003}
z^{l} F^{\prime}(\frac{1}{z}) -
{z^{l+1}}[\mu + \frac{l+2}{1-z}]F(\frac{1}{z}) =
- \frac{1}{\mu} z(1-z) F^{\prime \prime}(z)
- \frac{1}{\mu}(1 + zl) F^{\prime}(z)
+ \frac{1}{\mu z} F(z) -
\end{equation}
$$
- F(1)(l+2)\frac{z^{l+1}}{1-z}+
\frac{\psi(1)}{2(l+1)}
[(l+2-z) z^{l+1} + (-1)^l (1-z)^{l+2} +
\mu (-1)^l \frac{1 - z^{l+1}}{1-z}].
$$
The differential equation
$$
 - (z(1-z))^2 \frac{{\rm d}^3 F(z)}{{\rm d}z^3} +
(l - 1)z(1-z)(1-2z) \frac{{\rm d}^2 F(z)}{{\rm d}z^2} +
$$
\begin{equation}
+ [z(1-z)(l(l-1) - \mu^2 - \mu) + l+1+\mu^2 ]
\frac{{\rm d}F(z)}{{\rm d}z} -
\end{equation}
$$
- \mu^2 (l+2)\frac{1-2z}{z(1-z)} F(z) =  P(z),
$$
where
$$
P(z) = [(-1)^l(1-z)^{l+3} + z^{l+3}]
\mu(\mu - 1)(\mu + 2)\frac{\psi(1)}{2(l+1)} +
$$
$$
+[(-1)^l(1-z)^{l+2} + z^{l+2}]
(F(1)\mu^2(\mu + 1) +
 \frac{\psi(1)}{2(l+1)}
\mu(l+3 - \mu - \mu^2)) +
$$
$$
+ [(-1)^l z \frac{1-z^{l+1}}{1-z} +
 (1-z) \frac{1-(1-z)^{l+1}}{z}]\mu^3 \frac{\psi(1)}{2(l+1)}+
$$
$$
+ F(1)\mu^2(\mu+1) +
 \mu^3 \frac{\psi(1)}{2}[1 + (-1)^l] +
\mu^2 (l+2)F(1) \frac{z}{1-z}.
$$
is the condition of the compatibility of the
equations~(\ref{10002}) and~(\ref{10003}).

\begin{table}[h]
\label{Tabl1}
\caption{The table of eigenvalues }
\begin{tabular}{|l|l|l|l|l|l|}
\hline
L & $\mu_1$&  $\mu_3$&$\mu_2$ & $\mu^-_2$ & $\mu^+_2$   \\
\hline
L=1 &0.60000000&            &          &           &           \\
L=2 &0.71428571& 6.0000000  &          &           &           \\
L=3 &0.78857514& 4.0839167  &1.9759929 &           &           \\
L=4 &0.83937146& 3.5389047  &1.9503009 & -3.0418639&           \\
L=5 &0.87517807& 3.3100086  &1.9314066 & -3.0984570&           \\
L=6 &0.90102482& 3.1942326  &1.9199929 & -3.1590898& 4.0042568 \\
L=7 &0.92007929& 3.1288968  &1.9143589 & -3.2192912& 4.0132266 \\
L=8 &0.93440321& 3.0891924  &1.9126585 & -3.2770862& 4.0265764 \\
L=9 &0.94536854& 3.0637382  &1.9134658 & -3.3316057& 4.0436670 \\
L=10&0.95390533& 3.0467430  &1.9157926 & -3.3825103& 4.0637924 \\
L=15&0.97708364& 3.0129839  &1.9339767 & -3.5840909& 4.1893508 \\
L=20&0.98647363& 3.0048358  &1.9499351 & -3.7142619& 4.3229188 \\
L=25&0.99111713& 3.0021674  &1.9613325 & -3.7972717& 4.4433197 \\
L=30&0.99373325& 3.0011045  &1.9693761 & -3.8510501& 4.5446233 \\
L=40&0.99641060& 3.0003718  &1.9794625 & -3.9116771& 4.6926591 \\
L=50&0.99767975& 3.0001573  &1.9864961 & -3.9422393& 4.7860076 \\
L=60&0.99837854& 3.0000774  &1.9888526 & -3.9594671& 4.8451766 \\
L=70&0.99880357& 3.0000423  &1.9914543 & -3.9700495& 4.8838364 \\
\hline
\end{tabular}
\end{table}


\newpage
\begin{thebibliography}{99}
\bibitem{1}
S.K.Kehrein, F.J.Wegner and Y.M.Pis'mak,
Nucl. Phys. {\bf B402} (1993) 669
\bibitem{2}
S.K.Kehrein, F.J.Wegner,
Nucl. Phys. {\bf B424} (1994) 521
\bibitem{WK}
K.G.Wilson and J.Kogut,
Phys. Rep. {\bf 12} (1974) 75
\bibitem{YM}
Y.M.Makeenko,
Yad.Fiz., {\bf 33} (1981) 842
\bibitem{PD}
S.E.Derkachov and Yu.M.Pis'mak,
J.Phys.A: Math.Gen. {\bf 26} (1993) 1419
\bibitem{AN}
A.N.Vasil'ev, Sov. J. Teor. and Mat.Phys, v.{\bf 81}, n3, (1989) 336
\bibitem{Col}
D.C.Collins, "Renormalization", Cambridge Univ. Press, Cambridge, 1984.
\end{thebibliography}
\end{document}